\documentclass[3p,times]{elsarticle}

\usepackage{ecrc}

\usepackage{amssymb,amsmath,amscd}
\usepackage{graphicx,calc,epsfig,pstricks,bbm}
\usepackage{tikz}
\usepackage[T1]{fontenc}
\usepackage{ae,aecompl}
\usepackage{graphicx}	
\usepackage{amsmath}	
\usepackage[figuresright]{rotating}

\volume{00}
\firstpage{1}
\journalname{Physics Letters B}
\runauth{F. Cianfrani, G. Montani}

\jid{plb}

\begin{document}

\begin{frontmatter}

\title{Revised Conditions for MRI due to Isorotation Theorem}

\author[label1]{Francesco Cianfrani}
\author[label2]{Giovanni Montani}
\address[label1]{Institute
for Theoretical Physics, University of Wroc\l{}aw, Pl.\ Maksa Borna
9, Pl--50-204 Wroc\l{}aw, Poland.}
\address[label2]{ENEA, Fusion and Nuclear Safety Department, C.R. Frascati - Via E. Fermi, 45 (00044) Frascati (Roma), Italy\\
Physics Department, ``Sapienza'' University of Rome, P.le Aldo Moro 5, 00185 (Roma), Italy}

\begin{abstract}
We re-analyze the physical conditions for Magneto-rotational Instability (MRI) within a steady axisymmetric stratified disk of 
plasma, in order to account for the so-called 
isorotation theory (the spatial profile of differential angular velocity 
depends on the magnetic flux surface). 
We develop the study of linear stability around an 
astrophysical background configuration, following 
the original derivation in \cite{Ba:1995}, but implementing 
the isorotation condition as the orthogonality 
between the background magnetic field and the angular 
velocity gradient. 
We demonstrate that a dependence on the background 
magnetic field direction is restored in the dispersion 
relation and, hence, the emergence of MRI is also influenced by field orientation. 
\end{abstract}

\begin{keyword}
Magnetohydronamics (MHD) 


\end{keyword}

\end{frontmatter}




\section{Introduction and basic formalism}

The existence of an Alfv\'enic instability 
in weakly magnetized rotating plasmas (the plasma parameter $\beta$ is much greater than unity) 
was discovered in 1959 by E. P. Velikhov 
\cite{V59} and implemented on an astrophysical setting by S. Chandrasekar in 1960 \cite{C60} 
and it is well known as the Magneto-Rotational Instability (MRI). 

The intrinsic nature of such an unstable 
mode consists of the coupling 
between Alfv\'en waves and the plasma inhomogeneity, 
due to differential rotation (indeed MRI is suppressed in 
uniformly rotating systems). 
The most intriguing feature of MRI is, 
however, that it vanishes when the magnetic field has sufficiently high amplitude, 
because of the effect of the restoring 
force due to magnetic tension, which is
able to stabilize the plasma displacements. 

Despite being an instability driven by 
magnetic tension which does not transport matter, nonetheless MRI is very relevant 
for astrophysical systems and, in particular, 
for accretion structures. 

The Shakura idea for accretion \cite{S73,SS73}
relies on the existence of an effective 
viscosity, emerging from the plasma 
turbulence and the differential rotation of 
the accreting matter. 
Actually MRI turns out to be the most 
appropriate (if not the only reliable) 
mechanism to promote such a turbulent regime 
in astrophysical accreting systems
\cite{BH98,BH2003} (see also \cite{BH91}, 
where MRI was re-analyzed, after the 
original derivation). The great impact 
of this issue can be easily understood 
when realizing that most space plasmas are 
weakly magnetized, especially stellar 
disk configurations \cite{B01}. 

This scenario has led many authors, 
over the years, to investigate MRI in 
some different contexts and 
system morphologies, 
see for instance \cite{SM99,Ji:2001jp,Montani:2013uoa,Shakura:2014jda,Montani:2015rma,Montani:2014ppa}. 
A rather general result has been achieved 
in \cite{Ba:1995}, where a stratified 
and differentially rotating disk has 
been addressed in the context of MRI as a general property 
of axially symmetric profiles. 
The very surprising feature derived in \cite{Ba:1995}
consists of the independence of 
the stability conditions on the intensity 
and direction of the background magnetic 
field. The aim of the present analysis 
is to demonstrate how the direction 
of the magnetic field is actually 
involved in the stability of the stratified disk, as soon as the 
so-called isorotation theorem (also known as corotation theorem) is 
taken into account \cite{F37}. 
This theorem states that, for a steady, axially symmetric and purely rotating background (exactly the adopted hypotheses in \cite{Ba:1995}), the disk angular velocity 
depends on the magnetic flux function only. In what follows, this aspect will be equivalently formulated as the orthogonality 
character of the angular velocity 
gradient with respect to the background 
magnetic field. 

Here, we perform the linear perturbation analysis by 
accounting for the isorotation profile of the background plasma and 
then we demonstrate that the obtained dispersion 
relation takes the same form as 
the one in \cite{Ba:1995}, but after the orthogonality 
constraint between the background magnetic field and 
the angular velocity gradient is imposed. 
Then we investigate how such a revised dispersion relation 
affects the general stability properties of the stratified disk. 
The main merit of the present study consists of re-introducing 
a dependence of the stable profile on the angle that the 
background magnetic field forms with the entropy gradient, 
{\it i.e.} the extent of the stability region depends on the direction 
of the stationary magnetic configuration. 

A specific example is considered of a stratified disk around a compact massive astrophysical object, 
showing how a relevant stability region exists in the inner thick part of the disk. 

The basic theoretical framework of the present discussion is 
the ideal magnetohydrodynamic (MHD) 
theory, as viewed in axial symmetry, {\it i.e.} reducing the 
dynamical problem to a two-dimensional one.
In particular, in what follows, we will implement 
the momentum conservation equation: 

\begin{equation}
\rho \left( \partial _t\vec{v} + \vec{v}\cdot \vec{\nabla}\vec{v} 
\right) = -\vec{\nabla}p 
+ \frac{1}{4\pi}\left(\vec{\nabla}\wedge \vec{B}\right) \wedge \vec{B}
\, . 
\label{aw}
\end{equation}

and the induction equation: 

\begin{equation} 
\partial _t\vec{B} = \vec{\nabla} \wedge 
\left( \vec{v} \wedge \vec{B}\right) 
\, , 
\label{bw}
\end{equation} 

where $\vec{v}$ and $\vec{B}$ denote the velocity and 
magnetic fields, respectively, 
while $\rho$ is the plasma mass density and $p$ its total pressure. 

The ideal (hydrogen-like) plasma thermodynamics is regulated by the mass and 
entropy conservation equations, which read respectively 

\begin{align}
	\label{cw}
	&\partial _t\rho + \vec{\nabla}\cdot \left( 
\rho \vec{v}\right) = 0 
\\
&\left(\partial _t + \vec{v}\cdot \vec{\nabla}\right) 
\ln \left( p\rho^{-5/3}\right) = 0 
\label{dw} 
\, . 
\end{align}

Appropriately for the axial symmetry, 
in what follows we adopt standard 
cylindrical coordinates ($r,\phi ,z$) 
and we take both the background 
(denoted by a subscript $0$) and the 
perturbed (characterized by a corresponding subscript $1$) quantities 
as independent of the angle $\phi$, while the 
former are also stationary 
(independent of the time variable $t$).
More specifically, the background 
equilibrium is guaranteed by the simultaneous validity of the gravitostatic equilibrium and the Lorentz force-free condition, {\it i.e.} 
\begin{align}
\label{fw}
&\vec{\nabla} p_0 = \rho _0 
\left( \omega _0^2 r\hat{e}_r 
- \omega _K^2\hat{r}_p\right)  
\\
&\left( \vec{\nabla} \wedge \vec{B}_0 
\right) \wedge \vec{B}_0 = 0 
\, ,
\label{ew}
\end{align}

where $\omega _0$ is the background 
disk angular velocity, $\omega _K$ 
the Keplerian angular velocity and 
we adopt a compact notation 
for the poloidal vector radius 
({\it i.e.} $\hat{r}_p = (r,0,z)$). 
We consider separately the force-free 
condition, because we regard the background 
field $\vec{B}_0$ as the radiation one due to 
the central star, {\it i.e.} we assume, according to standard literature,
that, on the background level, the plasma does not provide a back-reaction field. Furthermore, the 
basic isorotation condition must hold \cite{F37}

\begin{equation}
\vec{B}_0\cdot \vec{\nabla}\omega _0 = 0\,.
\label{gw}
\end{equation}

The linear perturbation theory assumes the local approximation limit, 
in which the wavelength of the disturbances 
is much less than the typical scale of 
variation characterizing the background material, 
{\it i.e.} we will deal with the condition 
$\mid \vec{k}\mid \mid \vec{r}_p\mid \gg 1$
($\vec{k}$ being the wave vector used below)
. Such assumption allows to treat the 
background variables as parametric quantities and to use Fourier analysis ({\it i.e.} both in the radial and 
vertical directions the perturbations have short wave length compared to background profile variations).

It is worth noting that the implications of the isorotation theorem 
are relevant for a stratified thick disk only, since, in the case 
of a thin disk it holds automatically. In fact, for a thin 
disk, well confined near the equatorial plane, the magnetic field 
is essentially vertical, the angular velocity gradient is 
essentially radial, $\omega _0$ depending on the radial coordinate 
only, {\it and the vertical shear does not enter the problem}. Thus the present analysis, as that one in \cite{Ba:1995}, 
is physically relevant if the angular velocity changes rapidly enough 
along the vertical direction in comparison to the perturbation 
wavelengths (for discussion of perturbation analyses in the presence of vertical 
gradients see \cite{Liverts2010,Coppi2008}). The isorotation condition, stating the orthogonality 
between the background magnetic field and the angular velocity gradient, implies the relation

\begin{equation} 
\partial_z\omega_0 = - \frac{B_{0r}}{B_{0z}}\partial _r \omega _0 
\, . 
\label{xcxcc}
\end{equation} 

Thus the vertical shear is comparable to the radial one only in thick disk configurations 
(see for instance the dipole-like case of Section \ref{secIII}). 

Finally, we will adopt also the 
so-called Boussinesq approximation 
\cite{BH98}, which reduces the mass conservation 
equation to the local incompressibility 
of the perturbations, {\it i.e.} 
equation (\ref{cw}) becomes the 
constraint $\vec{k}\cdot \vec{v}_{p1} 
= 0$, where $\vec{v}_{p1}$ denotes the poloidal perturbed velocity. 
In a weakly magnetized plasma ($\beta\gg 1$), the sound speed 
$v_s$ is much larger than the Alfv\'en speed, since $\beta \sim 
v_s^2/v_A^2$. The limit of an arbitrarily large $v_s$ is 
equivalent to a diverging value of the polytropic index 
$\gamma$, associated to the equation of state 
$p_0 \propto \rho_0^{\gamma}$. In fact, we easily have 
$v_s^2 = \gamma p_0/\rho_0$ and the limit $\gamma \rightarrow 
\infty$ corresponds to an incompressible plasma 
(magnetized liquid), since a huge variation of the pressure 
corresponds to an infinitesimally small change in the mass 
density. Thus, for $\beta \gg 1$ the incompressibility condition, 
proper of the Boussinesq approximation, is adequate to 
describe the dominant features of the considered plasma system.

The study of the perturbation dynamics 
does not require, as we shall see, the 
specification of the equation of state, relating $p_1$ and $\rho_1$, since the perturbed pressure will be provided by requiring the 
validity of the local incompressibility 
during the plasma evolution. The 
resulting expression for $p_1$, for sufficiently 
large wavenumbers, results to be negligible 
in equation (\ref{dw}), which de facto becomes 
a link between $\rho _1$ and $\vec{v}_{p1}$.

\section{Linear approximation} 

Let us consider the case of axisymmetric perturbations of the form $(..)_1(t,\vec{r}_p)\propto e^{i(\vec{k}\cdot\vec{r}_p-\Omega t)}$, with a poloidal wave-vector $\vec{k}=(k_r,0,k_z)$ and frequency $\Omega$, according to the analysis in 
\cite{Ba:1995} that we are here revising (for a discussion on non axisymmetric perturbations allowed by the validity of the isorotation theorem see \cite{Montani:2014ppa}). 
The basic equations from which we start are the perturbed poloidal and azimuthal components of the momentum conservation equation \eqref{aw} and of the induction equation \eqref{bw} in the linear 
approximation, which read

\begin{align}
\Omega \vec{v}_{1p} - &2i\omega _0
v_{\phi_1 }
\vec{e}_r
-\frac{\vec{k}}{\rho _0}p_1 -i\frac{\vec{\nabla} p_0}
{\rho _0^2}\rho _1 -\frac{1}{4\pi \rho _0}\left(
\vec{B}_0\cdot \vec{B}_1\vec{k}
- \vec{k}\cdot \vec{B}_0\vec{B}_{1p}\right) = 0
\, 
\label{va}\\
\Omega v_{\phi_1} + &ir
\vec{v}_{1p}\cdot \vec{\nabla}
\omega _0 + 2i\omega _0v_{1pr}
+ \frac{\vec{k}\cdot \vec{B}_0}
{4\pi \rho _0}B_{\phi 1} = 0
\,
\label{va3}
\end{align}

and

\begin{align}
&\Omega \vec{B}_{1p} +
\vec{k}\cdot \vec{B}_0\vec{v}_{1p}=0
\label{va1}\\
&\Omega B_{\phi 1} -
i {r}\vec{B}_1\cdot \vec{\nabla}\omega _0
+ \vec{k}\cdot \vec{B}_0 v_{\phi_1} = 0
\,,
\label{va4}
\end{align}
respectively.

From \eqref{va1} and \eqref{va4} the components $\vec{B}_{1p}$ and $B_{\phi 1}$ of the magnetic field perturbation can be written in terms of other variables and the resulting expressions can be inserted into \eqref{va} \eqref{va3}. An expression for the perturbed pressure $p_1$ can be given by taking the scalar product of \eqref{va} with $\vec{k}$ and using the mass conservation equation in the Boussinesq approximation, namely $\vec{k}\cdot \vec{v}_{p1} = 0$, which must be preserved during the dynamics.
Eventually, the perturbed mass density can be given from the entropy conservation equation \eqref{dw}, which written for 
adiabatic perturbations, in the limit when $p_1/p_0\ll \rho _1/\rho_0$ (which is consistent with the large value of $k$ in the local 
approximation scheme, since $p_1$ is suppressed by a factor $\sim 1/k^2$), reduces to the form 

\begin{equation} 
i\Omega \frac{5}{3}\frac{\rho _1}{\rho _0} + 
\vec{v}_{1p}\cdot \vec{C}_0 = 0
\, , 
\label{va13}
\end{equation}

where $\vec{C}_0 \equiv \vec{\nabla} \ln 
\left( p_0\rho _0^{
-5/3}\right)$. 

The dispersion relation can be obtained by combining the conditions one gets from \eqref{va} and \eqref{va3} after all these manipulations and by considering the following decomposition for the wave vector: since, by virtue of the isorotation 
theorem \eqref{gw}, the two vectors $\vec{B}_0$ 
and $\vec{\nabla}\omega _0$ are orthogonal 
and the wavevector $\vec{k}$ has zero 
azimuthal component (the perturbations 
preserve axial symmetry), we can write

\begin{equation}
\vec{k} = k_\parallel \frac{\vec{B}_0}{\mid \vec{B}_0\mid} + k_\bot\frac{\vec{\nabla}\omega _0}{\mid \vec{\nabla}\omega _0\mid}
\, ,
\label{ort}
\end{equation}

or more explicitly

\begin{equation}
k_r = k_\parallel \cos \iota + k_\bot\sin \iota 
\, , \,  
k_z = -k_\parallel \sin \iota + k_\bot \cos \iota
\, , 
\label{kdef}
\end{equation}

with $0\le \iota < 2\pi$ being the angle between the background magnetic 
field and the equatorial plane, such that the background field reads $\vec{B}_0=|\vec{B}_0|(\cos\iota,0,-\sin\iota)$.
The axisymmetric configuration has a ``north-south'' symmetry, corresponding to the transformation $r \rightarrow r$ and $z\rightarrow -z$, 
under which the radial components of the magnetic field and the vertical components of vector fields (as the entropy gradient) change sign, while other components are unchanged. As a consequence, $\iota\rightarrow\pi-\iota$, $k_\parallel\rightarrow -k_\parallel$ and $k_\bot\rightarrow k_\bot$. In what follows, one can verify that the north-south symmetry is realized, since all the formulas contain invariant terms (as $k_\parallel^2\sin\iota$, $k_\parallel k_\bot \cos\iota$, ..).

The final dispersion relation is 
\begin{equation}
\tilde\Omega^6+A \tilde\Omega^4 + B \tilde{\Omega}^2 + C =0,\label{disprel}
\end{equation}
with
\begin{align}
A=&\frac{3}{5}\left(\frac{\vec{\nabla}p_0\cdot \vec{C}_0}{\rho _0} - \frac{(\vec{\nabla}p_0\cdot \vec{k})(\vec{k}\cdot\vec{C}_0)}{k^2\rho _0}\right)- y_m\frac{k_\parallel ^2}{k^2}\sin\iota+ y_m\frac{k_\parallel k_\bot}{k^2}\cos \iota  - 4\omega _0^2\frac{k_z^2}{k^2}\label{A} \\
B=&-\frac{3}{5}\left( y_m\frac{k_\parallel ^2}{k^2}\sin\iota - y_m\frac{k_\parallel k_\bot}{k^2}\cos \iota  + 4\omega _0^2\frac{k_z^2}{k^2}\right)\left(\frac{\vec{\nabla}p_0\cdot \vec{C}_0}{\rho _0} - \frac{(\vec{\nabla}p_0\cdot \vec{k})(\vec{k}\cdot\vec{C}_0)}{k^2\rho _0}
\right)-4\omega_A^2\omega _0^2\frac{k_z^2}{k^2}\nonumber\\
& + \frac{6}{5}\omega_0\, \Bigg[\left( \frac{r\vec{\nabla}p_0\cdot \vec{\nabla}\omega _0}{\rho _0} 
\frac{k_\parallel ^2}{k^2} - \frac{r\partial_\parallel p_0}{\rho _0}\mid \vec{\nabla} \omega
_0\mid \frac{k_\parallel k_\bot}{k^2} 
\right)+2\omega_0 \left(\frac{\partial _rp_0}{\rho _0}\frac{k_z^2}{k^2} - \frac{\partial _zp_0}{\rho
_0}\frac{k_rk_z}{k^2}\right)\Bigg]\nonumber\\
&\left( C_{0r}\frac{k_z^2}{k^2} - C_{0z}\frac{k_rk_z}{k^2} 
\right)\label{B}
\\
C=&\frac{12}{5}\omega_A^2\omega_0^2 \Bigg[-
\frac{k_z^2}{k^2}\left(\frac{\vec{\nabla}p_0\cdot \vec{C}_0}{\rho _0} - \frac{(\vec{\nabla}p_0\cdot \vec{k})(\vec{k}\cdot\vec{C}_0)}{k^2\rho _0}
\right)+\left( C_{0r}\frac{k_z^2}{k^2} - C_{0z}\frac{k_rk_z}{k^2} 
\right)\left( 
\frac{\partial _rp_0}{\rho _0}\frac{k_z^2}{k^2} - \frac{\partial _zp_0}{\rho
_0}\frac{k_rk_z}{k^2}\right) \Bigg]\,.
\end{align}

Above, we defined the quantities
$\omega ^2_A \equiv \frac{(\vec{k}\cdot\vec{B}_0)^2}{4\pi \rho _0}$, $\tilde{\Omega}^2=\Omega^2-\omega_A^2$ and $y_m=2\omega_0r|\vec{\nabla}\omega_0|$.

If as in \cite{Ba:1995} one introduces
\begin{equation}
\mathcal{D}=\frac{k_r}{k_z}\,\partial _z - \partial _r\,,
\end{equation}
it can be shown that
\begin{align}
&C_{0r}\frac{k_z^2}{k^2} - C_{0z}\frac{k_rk_z}{k^2}=-\frac{k_z^2}{k^2}\,\mathcal{D}\ln\left( p_0\rho _0^{-5/3}\right)\label{1}\\
&\frac{\vec{\nabla}p_0\cdot \vec{C}_0}{\rho _0} - \frac{(\vec{\nabla}p_0\cdot \vec{k})(\vec{k}\cdot\vec{C}_0)}{k^2\rho _0}=\frac{k_z^2}{k^2}\,\frac{\mathcal{D}p_0}{\rho _0}
\,\mathcal{D}\ln\left( p_0\rho _0^{-5/3}\right)\label{2}\\
&\frac{\partial _rp_0}{\rho _0}\frac{k_z^2}{k^2} - \frac{\partial _zp_0}{\rho
_0}\frac{k_rk_z}{k^2}=-\frac{k_z^2}{k^2}\,\frac{\mathcal{D}p_0}{\rho_0}\label{3}
\end{align}

from which it follows that $C=0$.  

Moreover, the following relation holds for the second term on the right-hand side of \eqref{A} 
\begin{equation}
- y_m\frac{k_\parallel ^2}{k^2} \sin\iota + y_m\frac{k_\parallel k_\bot}{k^2}\cos \iota  - 4\omega _0^2\frac{k_z^2}{k^2}=\frac{k_z^2}{k^2}\frac{1}{r^3}\,\mathcal{D}(\omega_0^2 r^4)\,. \label{relb}
\end{equation}
Similarly, using \eqref{1}, \eqref{2} and \eqref{relb} also the expression \eqref{B} for $B$ can be simplified. Collecting all these results, we end up with the following expressions
\begin{align}
A=&\left[\frac{3}{5}\frac{k^2_z}{k^2}\,\mathcal{D}\ln\left( p_0\rho _0^{-5/3}\right)\,\frac{\mathcal{D}p_0}{\rho_0}+\frac{k^2_z}{k^2}\,\frac{1}{r^3}\,\mathcal{D}(\omega_0^2r^4)\right]\Bigg|_{cr}\\
B=&\left[-  4\omega_A^2\omega _0^2\frac{k_z^2}{k^2}\right]\Bigg|_{cr}\\
C=&0\,,
\end{align}
where $|_{cr}$ means the imposition of the isorotation theorem, {\it i.e.} the decomposition \eqref{ort} for $\vec{k}$. The resulting dispersion relation formally coincides with that given by Balbus in \cite{Ba:1995}, but for the implementation of the isorotation theorem. 
By other words, we are demonstrating that the isorotation theorem can be also imposed after the dispersion relation has been derived. Nonetheless, this leads to significant modification of the stability conditions, which we are going to analyze in detail.
 
The study of the stability can be performed as in \cite{Ba:1995}. As it can be realized from the dispersion relation, $\Omega^2$ is real and we have stability for $\Omega^2>0$ and instability for $\Omega^2<0$. Since, there always exists a stable solution with $\Omega^2=\omega_A^2 $,  the stability condition can be translated into the condition that the dispersion relation \eqref{disprel} does not admit solutions in the limit $\Omega\rightarrow 0$, which implies the following inequality 
\begin{align}
&x^2\,N_z^2+x\,\Big[\frac{3}{5\rho_0}\bigg(\partial_zp_0\,\partial_r\ln(p_0\rho_0^{-5/3})+\partial_rp_0\,\partial_z\ln(p_0\rho_0^{-5/3})\bigg)-r\,\partial_z\omega_0^2\Big]+N_r^2+r\,\partial_r\omega_0^2>0\,, 
\end{align}
where
\begin{equation}
x=\frac{k_r}{k_z}\,,\qquad N^2_i=-\frac{3}{5\rho_0}\,\partial_ip_0\,\partial_i\ln(p_0\rho_0^{-5/3})\,.
\end{equation}
The inequality above is always verified if (we note that in \cite{Ba:1995} the first stability condition is different, but the whole set of conditions can be easily shown to be equivalent) 
\begin{align}
&N_z^2>0\label{stab1}\\ 
-\partial_z p_0&\,\left[\partial_r \omega^2_0\,\partial_z \ln\left( p_0\rho _0^{-5/3}\right) - \partial_z \omega^2_0 \, \partial_r \ln\left( p_0\rho _0^{-5/3}\right) \right]>0\,.\label{stab2}
\end{align}

The imposition of the isorotation theorem here means that the derivatives of $\omega_0$ must be written according to the following relations
\begin{equation}
r\,\partial_r \omega_0^2= y_m\,\sin\iota\qquad r\,\partial_z \omega_0^2= y_m\,\cos\iota\,,
\end{equation}
which come from the decomposition of the radial and vertical directions as in \eqref{ort}. Therefore, the stability condition \eqref{stab2} becomes 
\begin{equation}
-y_m\,\partial_z p_0\,\left[\sin\iota\,\partial_z \ln\left( p_0\rho _0^{-5/3}\right) - \cos\iota \, \partial_r \ln\left( p_0\rho _0^{-5/3}\right) \right]>0\,,\label{sc2}
\end{equation}
which can be rewritten as follows 
\begin{equation}
-y_m\,\partial_z p_0\,\left(\vec{B}_0\cdot\vec{C}_0\right)>0\,.\label{sc3}
\end{equation}
This relation outlines how for a given matter configuration, {\it i.e.} given the spatial profiles of $\rho_0$ and $p_0$, the stability of the system depends on the direction of the magnetic field. It is worth noting how the north-south symmetry is still realized, since for $z\rightarrow -z$ both $\partial_z p_0$ and $\vec{B}_0\cdot\vec{C}_0$ change sign ($\vec{B}_0$ and $C_0$ are pseudo-vector and vector fields, respectively), such that \eqref{sc3} is unchanged.  

Summarizing, we demonstrated that deriving the dispersion relation when the 
isorotation theorem holds is equivalent to directly impose such a restriction 
on the spectrum presented in \cite{Ba:1995}. By other words, we revised the analysis 
of Balbus, by recognizing that the disk background must satisfy the 
isorotation condition and hence the dependence of the dispersion relation on 
the background magnetic field direction is revealed, under the assumption of isorotation.

It is worth noting that a similar conclusion can 
be inferred also from the global analysis of \cite{Papaloizou(1992)}. The dependence on the magnetic field direction 
is there inside the adopted coordinates $\psi$ and $\chi$, which are orthogonal and parallel to $\vec{B}_0$, respectively. 
Hence, any dependence on these variables is indeed a dependence on the magnetic field direction. For instance, the stability condition 
(38) of \cite{Papaloizou(1992)} can be rewritten in terms of the scalar product between the magnetic field and the entropy gradient, just like we did in \eqref{sc3}, due to the relation
\begin{equation}
\frac{\partial_\chi p_0}{\Gamma p_0} -\frac{\partial_\chi \rho}{\rho}\propto \vec{B}_0\cdot \vec{C}_0\,.  
\end{equation}

We end this section by giving the explicit form of the growth rate $\gamma$, for which $\gamma^2=-\Omega^2>0$. Let us first introduce the following quantities 
\begin{equation}
\omega_t^2= \frac{\mathcal{D}p_0}{\rho _0}
\,\mathcal{D}\ln\left( p_0\rho _0^{-5/3}\right)\qquad \omega^2_e=\frac{1}{r^3}\mathcal{D}\left(\omega_0^2 r^4\right)\qquad v_A^2=\frac{\omega^2_A}{k^2}\,,
\end{equation}
the growth rate is
\begin{equation}
\gamma^2=-k^2\,v_A^2+\frac{3}{10}\frac{k_z^2}{k^2}(\omega_t^2+\omega^2_e)+\frac{1}{2}\sqrt{\frac{9}{25}\frac{k_z^4}{k^4}(\omega_t^2+\omega^2_e)^2+16\,v_A^2\,\omega_0^2\,k^2}\,,
\end{equation} 
where the admissible values of $k^2$ are those for which $\gamma^2>0$. As a function of $k^2$, $\gamma$ has a maximum in $k^2=\frac{\omega_0^2}{v_A^2}-\frac{9}{400}\frac{k_z^4}{k^4}\frac{(\omega_t^2+\omega_e^2)^2}{\omega_0^2 v_A^2}$ corresponding to  
\begin{equation}
\gamma^2_{max}=\omega_0^2+\frac{3}{10}\frac{k_z^2}{k^2}(\omega_t^2+\omega^2_e) + \frac{9}{400} \frac{k_z^4}{k^4}\frac{(\omega_t^2+\omega^2_e)^2}{\omega_0^2}\,.
\end{equation}

\section{Relevant application}\label{secIII}

Let us consider as an application of the previous analysis the case of a specific background morphology, describing a stratified (thick) accretion disk around a compact astrophysical object. According to equations \eqref{fw}, the background magnetic field is essentially that generated by the central object and it is here modeled by a dipole-like configuration (a reliable approximation for many concrete astrophysical systems), {\it i.e.} it is determined by the function $\psi=\frac{N r^2}{(r^2+z^2)^{3/2}} $, where the constant $N$ fixes the magnetic field amplitude, and it leads to the following magnetic fields components
\begin{equation}
\vec{B}= \vec{\nabla} \wedge \left(\frac{\psi}{r}\right)\, \hat{e}_{\phi}=\frac{N}{(r^2+z^2)^{5/2}}(3rz,0,2z^2-r^2)\,.
\end{equation}
The gradients of the pressure can be inferred from \eqref{fw}, by assuming $\omega^2_0=GM\,r^6/(r^2+z^2)^{9/2}$, $M$ being the mass of the central object. This assumption comes from implementing the isorotation theorem and requiring that on the equatorial plane the angular velocity be Keplerian \cite{Montani:2012gq}. We investigate the equations for the pressure gradients \eqref{fw} and implement the consistency condition for which a solution exists (namely, that the mixed derivatives of the pressure coincide). This condition provides a restriction on the admissible form of the energy density: it turns out that the matter density is the product of $(r^2+z^2)^{9/2}$ times an arbitrary function of $r^8-(r^2+z^2)^4$, while the pressure can be explicitly computed and it equals the primitive of such arbitrary function modulo a constant. We stress that this form of the energy density and pressure is a mathematical requirement for the existence of a pressure profile for which \eqref{fw} holds. The choice of the arbitrary function of $r^8-(r^2+z^2)^4$ determines the physical properties of the considered matter distribution, but any choice is admissible mathematically. We choose an exponential function, leading to the following particular configuration
\begin{equation}
\rho_0=K\,(r^2+z^2)^{9/2}\,e^{\frac{r^8-(r^2+z^2)^4}{\ell^8}}\quad p_0=K\ell^8 \,e^{\frac{r^8-(r^2+z^2)^4}{\ell^8}}\,,
\end{equation}
$K$ and $\ell$ being two integration constants. It is easy to check that these expressions solve \eqref{fw} by direct substitution. The corresponding profile is characterized by an exponential decay along $z$, with a decreasing decay length along the radial direction (see figure 1
).

\begin{figure}[htb!]
	\centering
		\includegraphics[width=0.40\textwidth]{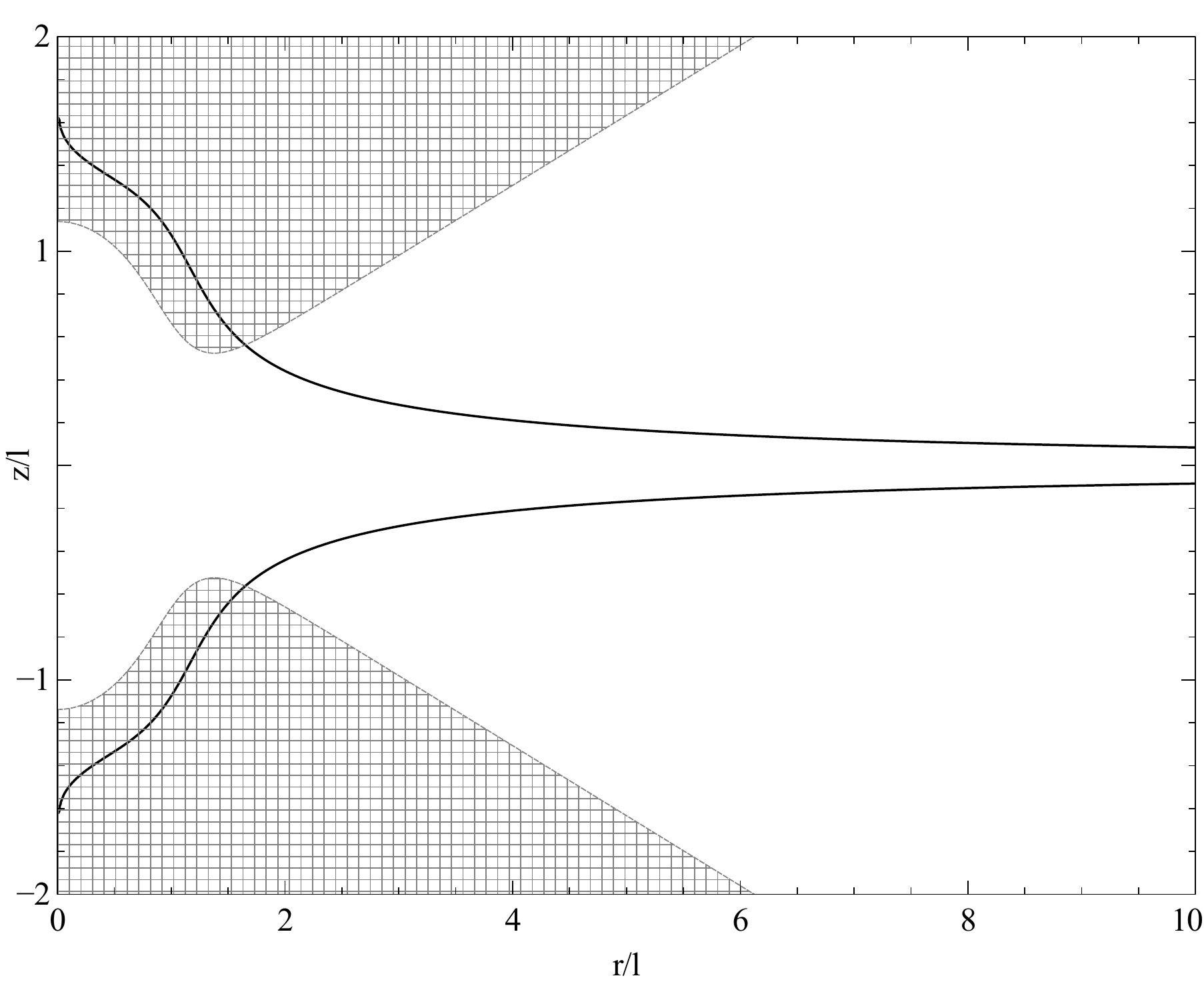}
	\label{fig:graficodisco}
	\caption{The stable region of the system, denoted by the crossed area, is compared with its decay length (defined as the distance where the mass density is reduced by a factor $e$ with respect to the equatorial value), represented by the solid line. It is worth noting how the stable region comprises some areas in which the matter density is not negligible in the thick region of the disk.}
\end{figure}

The stability condition \eqref{stab1} can be shown to be already contained into \eqref{sc2}, such that the stable region is solely determined by the direction of the magnetic field according to \eqref{sc3} (it is drawn in figure 1
). This is clarified in figure 2
: in the stable region the angle between the magnetic field and the entropy gradient is acute, while in the unstable region it is obtuse.   

This analysis outlines how the relative orientation of the magnetic field with respect to the entropy gradient is crucial for the stability of the whole system. In particular, for the considered configuration MRI is suppressed close to the central object in the inner part of the stratified disk (in correspondence to the lobes in figure 1) and therefore no turbulent transport can easily emerge there. 


\begin{figure}[htb!]
	\includegraphics[width=0.40\textwidth]{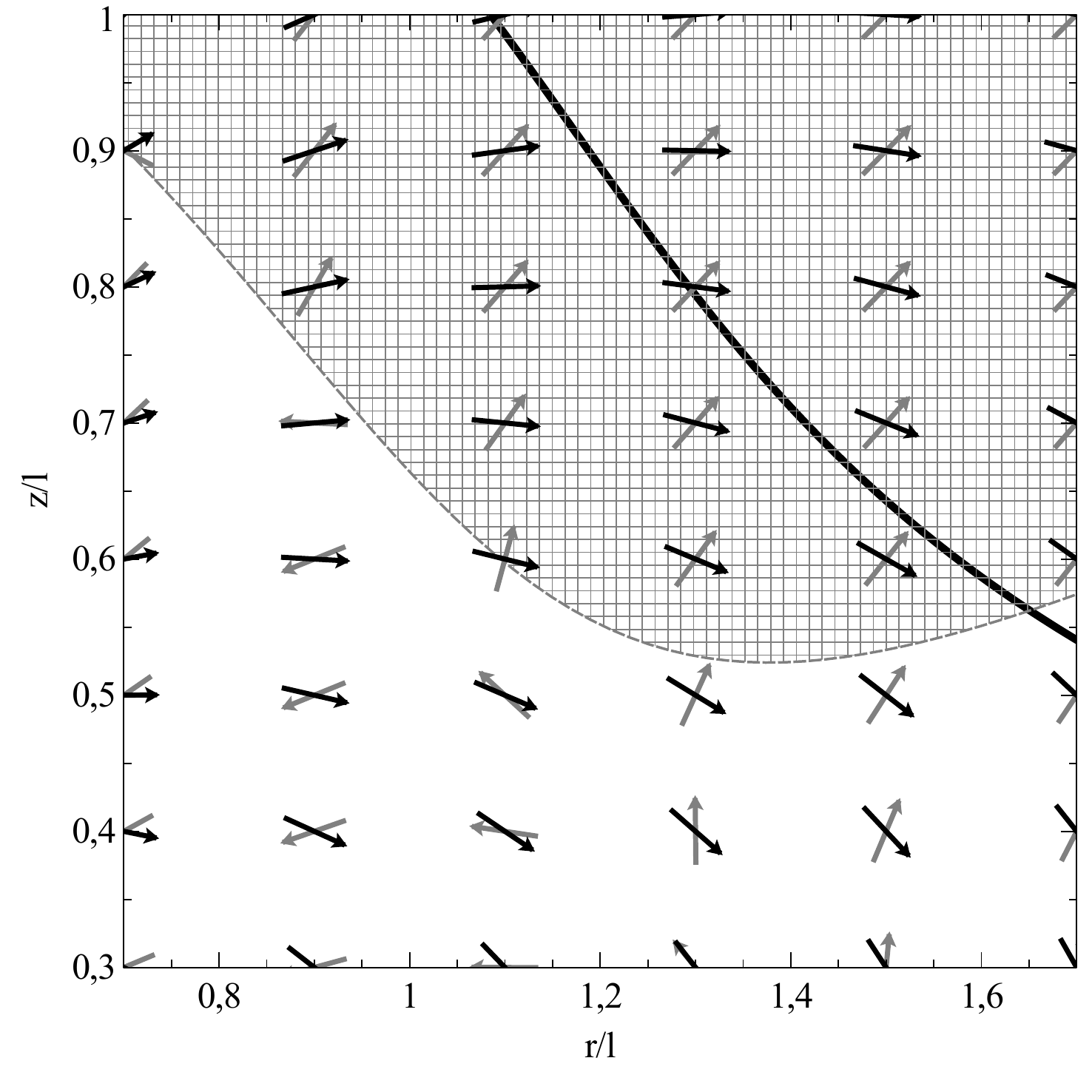}
	\label{fig:vector}
		\caption{The magnetic field and the entropy gradient, denoted by black and grey vectors, respectively, are evaluated near the boundary between the stable and unstable regions. In particular, the angle between them is acute/obtuse in the stable/unstable region.}
\end{figure}

We observe how, far enough from the symmetry centre, the 
disk configuration is essentially thin and the stability 
region would take place in the outer part of the disk, 
{\it i.e.} at a bigger vertical distance than the characteristic 
decay length of the configuration (defined as the distance at which the density 
is reduced by a factor $e$ with respect to its value at the equator).
This means that the stability region is of little relevance far from the centre.

The relevant modification of the spectrum with respect to \cite{Ba:1995} takes  
place close to the centre of the disk configuration, where the stability domain overlaps 
the region in which the vertical distance is smaller than the decay length. This means that 
the stability condition holds in a domain where the disk mass density is significant.

We conclude this section observing how the provided example describes an accretion disk very far from a thin disk
model: in the inner region the configuration has a spherical profile and its mass density rapidly decays outward. 
Such a thick and stratified configuration is however predicted by the equilibrium and, in this respect, it does not 
constitute a limiting case. In Nature, plasma disks of this sort are, for instance, the so-called Advective 
Dominated Accretion Flows \cite{Narayan(2008)}, or tori surrounding Active Galactic Nuclei. 

Indeed, the dependence we traced above of the stability on the magnetic field direction better emerges in
configurations allowing a significant range of variation for the ratio of the poloidal magnetic components.

\section{Concluding remarks}

We developed a perturbation analysis which accounts 
for the validity of the isorotation theorem on the 
background disk configuration, here translated as the 
orthogonality of the background magnetic field and 
the angular velocity gradients. The effect of such a 
restriction on the dispersion relation corresponds to 
restore a dependence on the background magnetic field 
direction into the mode spectrum (absent in \cite{Ba:1995}). 
Such an effect is relevant only for a stratified 
plasma disk since, when the $z$-dependence is 
weak or removed (like in a thin disk morphology), 
the isorotation constraint is automatically satisfied: 
the magnetic field is mainly vertical and the 
angular velocity gradient is essentially radially directed. 

We have seen how, in a specific sample, the restored magnetic field dependence 
leads to a significant modification of the MRI morphology 
across the disk, since a threshold angle emerges. 
In fact, MRI can develop only when the angle between the magnetic field and the entropy gradient of the background 
configuration is bigger than $\pi/2$.

From an astrophysical point of view, the 
obtained result could affect the transport properties 
of thick stratified disks, since in some regions 
of the plasma profile the turbulent flow due to 
MRI is clearly suppressed (as elucidated in the previous section), together with the 
associated effective viscosity. 
This feature is particularly enhanced far from the 
equatorial plane (when also the dipole magnetic field 
is significantly inclined) and for stiff radial pressure gradients, 
{\it i.e.} when the disk rotation deviates from the 
Keplerian behavior \cite{Ogilvie}. 

Independently from the astrophysical range of applicability 
for the isorotation theorem (for a discussion see 
\cite{Montani:2014ppa}), the present study removes 
the surprising and, a bit unnatural, feature discussed 
in \cite{Ba:1995}, about the independence of MRI spectrum 
on the magnetic configuration, {\it i.e.} on its own trigger 
mechanism: actually the magnetic field direction enters the 
spectrum and also a forbidden spatial region for MRI 
emerges. 

\section*{Acknowledgements}

This work has been developed in the framework of the CGW Collaboration
(www.cgwcollaboration.it).

\end{document}